\documentclass[a4page,10pt]{article}
\usepackage[top=0.8in, bottom=0.8in, left=0.5in, right=0.5in]{geometry}
\usepackage{fancyhdr}
\renewcommand{\thispagestyle}[1]{}
\usepackage[pdftex]{graphicx}
\usepackage[utf8]{inputenc}
\usepackage{hyperref}
\usepackage{tabularx}
\usepackage{floatrow}
\usepackage{multirow}
\newfloatcommand{capbtabbox}{table}[][\FBwidth]

\title{Minimally-Supervised Attribute Fusion for Data Lakes}

\begin{document}
\author{Karamjit Singh, Garima Gupta, Gautam Shroff, and Puneet Agarwal
%
%
\vspace{.3cm}\\
%
TCS Research, New-Delhi, India\\
%
}
\maketitle

\begin{abstract}
Aggregate analysis, such as comparing country-wise sales versus global market share across product categories, is often complicated by the unavailability of common join attributes, e.g., category, across diverse datasets from different geographies
or retail chains, even after disparate data is technically ingested into a common \textit{data lake}.
Sometimes this is a missing data issue, while in other cases it may be inherent, e.g., the records in different 
geographical databases may actually describe different product `SKUs', or follow different norms
for categorization. Record linkage techniques, such as \cite{christen2008febrl} can be used to automatically map products in different data sources
to a common set of \textit{global} attributes, thereby enabling federated aggregation joins to be performed.
Traditional record-linkage techniques are typically unsupervised, relying textual similarity features across attributes
to estimate matches. In this paper, we present an ensemble model combining minimal supervision using Bayesian network
models together with unsupervised textual matching for automating such `attribute fusion'. We present results of our approach 
on a large volume of real-life data from a market-research scenario and compare with a standard record matching algorithm. 
Finally we illustrate how attribute fusion using machine 
learning could be included as a data-lake management feature, especially as our approach also
provides confidence values for matches, enabling human intervention, if required.
\end{abstract}

\section{Introduction}
Traditional business intelligence is rapidly evolving to adopt modern big-data analytics architectures based on the concept of a `data lake', where, rather than  first integrating multiple historical data from diverse sources into a common star schema via extraction-transformation-load operations, the datasets are maintained in their raw form. This leads to a number of challenges; for example, dealing with incongruous join keys between different datasets. 

In this paper, we focus on a problem of fusion of information about consumer products, such as sales, market share, etc., which is spread across disparate databases belonging to different organizations, across which a product is \textit{not} identifiable via a common key. For example, a \textit{Global database (DB)}
might track overall market-share of global product categories. On the other hand, each \textit{Local DB} might track sales data within a geography 
using local-product-ids along with other characteristics, but \textit{not} the global category-id. As a result, an analytical task such as comparing the
sales of product categories within each geograpy against their global market share becomes difficult due to the lack of a natural join attribute between the databases.
(Note that the same product might be characterized using different attributes in different countries, including also textual \textit{description} of products entered manually by retailers, e.g., for carbonated drinks it usually contains information of brand, size, material used, packaging etc.) 


One way to perform analysis across disparate databases is by mapping records in each \textit{Local DB} to their corresponding global attributes (e.g., `category'
in the example above). However, preparing such mappings is a huge manual and complicated task because: a)~The cardinality (number of possible values) of local and global characteristics varies from tens to thousands, and b)~Uncertainty in the semantics of local characteristics of the same product from different geographies, leading to confusion in identifying the product category, even by human annotators.
 
Our aim is to help reduce cost of the operational process of creating and maintaining such global references by reducing manual workload via automation
via modern data-lake architecture that include automated fusion of federated databases. Our goal is to either make high confidence predictions, or abstain from making any prediction so that such records can be sent to human annotators. We want to minimize the number of such abstentions while maximizing the precision of the predictions.

\begin{figure}
\centering
\includegraphics[width=120mm]{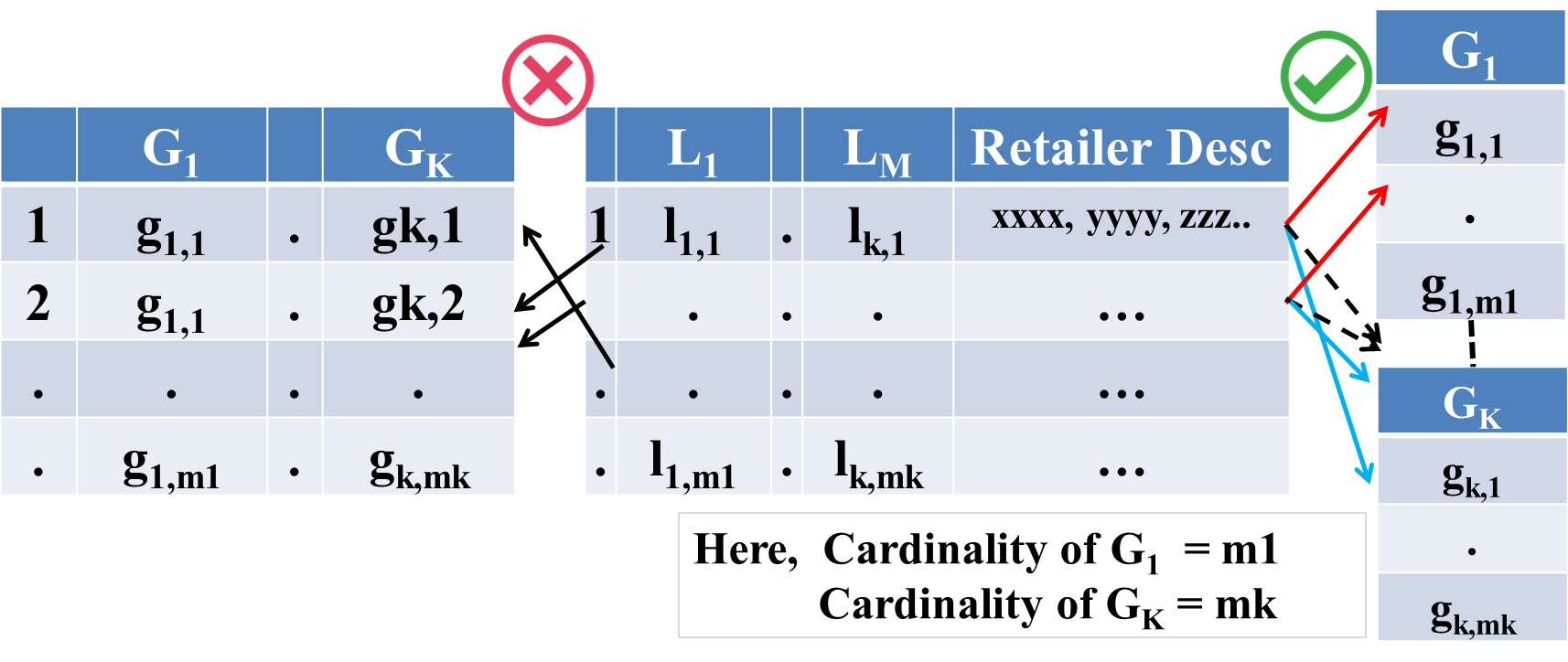}
\caption{Local and Global Database}
\label{fig:Prob1}
\end{figure}

\textbf{\textit{Attribute Fusion using Record Matching}}:
Consider two databases (see Figure~\ref{fig:Prob1}): a) \textit{Local DB($L$)} with each product $l$ having local characteristics $L_1, L_2,..., L_M$, e.g., flavor, brand, etc., and retailer descriptions ($D_i$), and b)~a \textit{Global DB($G$)} having $K$ global characteristics. The problem at hand is thus a \textit{record matching} problem where products in local database are to be mapped to global characteristic values (e.g. `category', or `global brand' etc.). 

Note that our objective is only to reconcile performance metrics (such as volume sales and market-share) across databases for each global characteristic \textit{independently}, e.g., sales vs market-share for each category, or alternatively each global brand, etc.
We can achieve this by solving $K$ different record matching problems,
as shown in Figure~\ref{fig:Prob1}: For each product, we shall predict each of the $K$ global characteristics
given local characteristics and retailer descriptions separately, as $\arg\max_j P(G_j | L_1, ..., L_M, D_i)$.

In this paper: a)~We address the problem of automating attribute fusion across diverse data sources that do not share a common join key.
b)~We augment traditional, fundamentally unsupervised text-similarly techniques with supervised, Bayesian network models in a confidence-based ensemble for automating the mapping process.
c)~Our approach additionally delivers confidence bounds on its predictions, so that human annotation can be employed when needed.
d)~We test our approach in a real-life market research scenario. We also compare it with available techniques \cite{christen2008febrl} and demonstrate that our approach outperforms FEBRL \cite{christen2008febrl}.
e) We illustrate how our approach has been integrated into a data-fusion platform\cite{singh2016visual} specifically designed to manage data-lakes containing disparate databases.

\textbf{Related Work:} Record linkage has been usually addressed via two categories of approaches, learning-based and non-learning based~\cite{kopcke2010evaluation}. Learning-based approaches such as FEBRL\cite{christen2008febrl} that uses SVM to learn a weighted combination
of similarity matching techniques followed by unsupervised matching, MARLIN~\cite{bilenko2003adaptive} uses similarity measures Edit Distance and Cosine and several learners. In non-learning based approaches, PPJoin+\cite{xiao2011efficient} is a single-attribute match approach using sophisticated filtering techniques for improved efficiency, and FellegiSunter\cite{fellegi1969theory} evaluates three of the similarity measures Winkler, Tokenset, Trigram. In\cite{poon2016ensemble}, an ensemble approach of two non-learning algorithms Fellegi-Sunter and Jaro-Wrinkler has been presented for record-linkage. In contrast, we use a confidence based ensemble approach that combines supervised learning using a Bayesian network model together with a non-learning based textual model.
Our approach also produces confidence bounds on the predictions that help to decide reliability of prediction.

\label{sec:intro}
\section{Approach}
Each product $l$ in $L$ has two kind of information (1) $M$ Local characteristics and (2) Textual descriptions by retailers. In this section, we present our approach to predict the value of global characteristic $G_j$ for each product in $L$. We use two different models for two different datasets (1) Supervised Bayesian Model (SBM) using local characteristics, and (2) Unsupervised Textual Similarity (UTS) using descriptions to compute probability of every possible state $g_{j,t}, t = 1, 2, ..., m j$ of $G_j$. Finally, we use an weighted ensemble based approach to combine the probabilities of both models to predict the value of $G_j$.
\subsection{Supervised Bayesian Model}\label{sec:BGM}
Approach to build SBM comprises of:(1)~Network Structure Learning, (2)~Parameter Learning, \& (3)~Bayesian Inference. For structure learning, we propose a novel technique of learning Tree based Bayesian Networks(TBN), whereas for parameter learning and Bayesian inference, we use the idea of \cite{yadav2015business} that performs probabilistic queries using SQL queries on the database of conditional probability tables.

\textbf{TBN Structure Learning:}\label{sec:struct}
Bayesian networks are associated with parameters known as conditional probability tables (CPT), where a CPT of a node indicates the probability that each value of a node can take given all combinations of values of its parent nodes. In CPTs, the number of bins grows exponentially as the number of parents increases leaving fewer data instances in each bin for estimating the parameters. Thus, sparser structures often provide better estimation of the underlying distribution~\cite{koller2009probabilistic}. Also, if the number of states of each node becomes high and the learned model is complex, Bayesian inferencing becomes conceptually and computationally intractable~\cite{lam1994learning}.
Hence, tree-based structures can be useful for density estimation from limited data and in the presence of higher number of states for facilitating faster inferencing. We employ a greedy search, and score based approach for learning TBN structure.

Given the global characteristic $G_j$ and $M$ local characteristics, we find set of top $\eta$ most relevant local characteristics w.r.t. $G_j$ using mutual information. We denote these $\eta$ local characteristics by $Y^j(L)$. Further, we learn a \textit{Tree based Bayesian Network(TBN)} on random variables $X = \{X_r: r=1,2,...,\eta+1\}$, where each $X_r \in X$ is either local characteristic $L_i \in Y^j(L)$ or global characteristic $G_j$ 

Chow et al. in \cite{chow1968approximating} state that cross-entropy between the tree structures distributions and the actual underlying distribution is minimized when the structure is a maximum weight spanning tree(MST). So, in order to learn TBN structure, we first learn MST for the characteristics in the set $X$. We find the mutual information between each pair characteristics, denoted by $W(X_r,X_s)$. Further, we use the mutual information as the weight between each pair of characteristics and learn MST using Kruskal's algorithm.

\vspace{-15pt}
\begin{equation}
\scriptsize
\label{eq:cross_ent}
Total Weight(TW) = \sum_{r=1,Pa(X_{r}) \neq 0}^{\eta+1} W(X_{r},Pa(X_{r}))
\vspace{-5pt}
\end{equation}

By learning MST, order of search space of possible graphs is reduced to $2^{O(\eta)}$, from $2^{O((\eta)^2)}$. Using this MST we search for the directed graph with least cross-entropy, by flipping each edge directions sequentially to obtain $2^{\eta}$ directed graphs along with their corresponding TW calculated using Eq.~\ref{eq:cross_ent}. Graph with maximum TW (minimum cross-entropy)~\cite{lam1994learning} is chosen as the best graphical structure representative of underlying distribution.

\textbf{Parameter Learning and Inference:} To learn the parameters of Bayesian Network(CPTs), for every product $l$ in $L$ we compute the probabilities $p^l_{j,1}, p^l_{j,2},..., p^l_{j,m_j}$, for every state of $G_j$, given the observed values of local characteristics in the Bayesian network, using an approach described in\cite{yadav2015business}. Here, CPTs are learned from the data stored in RDBMS and all queries are also answered using SQL.
\subsection{Unsupervised Text Similarity}\label{sec:tir}
In this section, we present UTS approach to compute the probability $q^{l}_{j,t}$ of each possible state of the global characteristic $G_j$ using retailer descriptions.
Consider each product $l$ in $L$ has $r_l$ descriptions and 
for each description $d_{l,r}$, where $r=1,2,..., r_l$, we find n-grams of adjacent words. Let $N_{l} = \{ n^l_{v}, v=1,2,...\}$ be the set of n-grams of all descriptions, where $f^l_{v}$ be the frequency of each $n^l_{v}$ defined as a ratio of the number of descriptions in which $n^l_{v}$ exists to the $r_l$.

For every state $g_{j,t}$ of $G_j$, we find the best matching n-gram from the set $N_l$ by calculating Jaro-Wrinkler distance between $g_{j,t}$ and every $n^l_v \in N_l$ and choose the n-gram, say $n^l_{v,t}$, with the maximum score $s^l_{j,t}$.  
Further, multiply the scores $s^l_{j,t}$ with the frequency of $n^l_{v,t}$ to get the new score i.e., $S^{l}_{j,t} = s^{l}_{j,t} \times f^s_{l,t}$.
Finally, we convert each score $S^{l}_{j,t}$ into the probability $q^{l}_{j,t}$ by using softmax scaling function.

\subsection{Ensemble of models}\label{sec:ensemble}
In ensemble approach, we first find confidence of each prediction in both the cases(SBM and UTS) and then use these confidence values as weights for weighted ensemble. Given the probability distribution \{$p^l_{j,t}: t=1, 2,..., m_j\}$ for the values of $G_j$ using SBM model, we find the confidence corresponds to each probability as
\vspace{-7pt}
\begin{equation}
\scriptsize
 C(p^l_{j,t}) = 1 -  \sqrt{\sum_{t^{'}=1}^{m_j}(p^l_{j,t^{'}} - h^l_{t^{'}}(t))^2}, t=1,2...,m_j
 \vspace{-8pt}
\label{eq:conf}
\end{equation}
where $h^l_{t^{'}}(t)$ is the ideal distribution, which is 1 when $t = t^{'}$ and 0 otherwise.
Similarly, we can find the confidence $C(q^l_{j,t})$ of each probability $q^l_{j,t}$.

With the given probability dist. and the confidence values from both models, we take weighted linear sum of two probabilities to get the new probability distribution over the states of {\scriptsize $G_j$: $P^l_{j,t} = C(p^l_{j,t}) \times p^l_{j,t} + C(q^l_{j,t}) \times q^l_{j,t}$, $t = 1,2,..., m_j$}
and we choose the value of $G_j$ for maximum $P^l_{j,t}$.

\textbf{CoP}: For every prediction, we assign the confidence value called confidence of prediction (CoP). CoP is a measure that helps to decide whether the predicted value is trustworthy or not. Given the probability distribution \{$P^l_{j,t}: l=1,2,...,m_j$\} for the values of $g_j$, we calculate the CoP of the predicted value $g^l_{j,t}$ of $G_j$ by using Equation~\ref{eq:conf}.

\section{Experiments and results}\label{sec:experiment}
\begin{figure}
\centering
\includegraphics[width = 100mm]{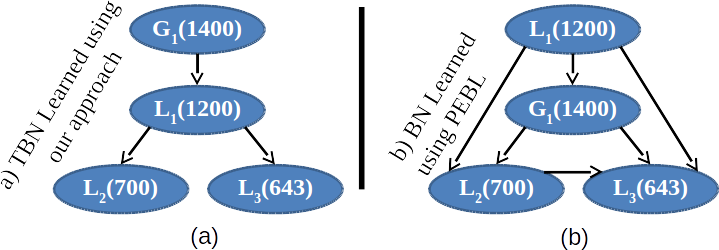}
\vspace{6pt}
\caption{Structure learned of $G_1$ using two different approaches}
\label{fig:g1net}
\end{figure}

We present the accuracy of our predictions on a real-life dataset from a global market research organization. We set a threshold $\tau$ on CoP, and predictions with a CoP $< \tau$ are routed for human annotation. We also measure the accuracies of our predictions for different values of $\tau$.

\textit{\textbf{Data description}}: We have data for carbonated drinks of $26$K unique products from a single geography, contained in two datasets:
a)~\textbf{Local DB}: It contains 26K products with each product having $49$ local characteristics, where cardinality of local characteristics varies from tens to thousands. It also contains descriptions of products given by retailers of that product, where number of descriptions of a single product varies from tens to hundreds. b)~\textbf{Global DB}: It contains four global characteristics with cardinality varying from tens to thousands.

\textbf{\textit{Data Preparation:}} We predict four global characteristics $G_1$, $G_2$, $G_3$, and $G_4$ for two cases, with varying ratio of split between training, validation and test datasets. \textbf{\textit{Case-1}}(60:20:20) has 60\% training, 20\% validation and 20\% test and \textbf{\textit{Case-2}} has this ratio as 20:20:60. \textbf{NOTE:} While Case-1 uses a traditional split of training vs testing data, Case-2 is more realistic, since in practice preparing a training data by manual data labeling is costly: For example, we would like to `onboard' a data from a particular dataset 
by manually annotating only a small fraction (e.g. 20\%) of records and automate the remainder or we might like to board 
data from one organization (e.g. retailer or distributor) in a particular geography in the hope that data from remaining sources in that
geography share similar local characteristics, eliminating manual annotation for a large volume of data. 
To simulate this practical scenario, we used the first few records from the local dataset, which happened to contain only 10\% or so of the
total possible values of each global attribute.

For SBM, $\eta$ relevant local characteristics was chosen for every $G_j$. Figure~\ref{fig:g1net}, compares the Bayesian network structure learned using our approach and another learned using an open source python library Pebl \cite{shah2009python}, for the global characteristic $G_1$. Clearly, network obtained using Pebl (Figure~\ref{fig:g1net}(b)) is more complex as compared to ours \ref{fig:g1net}(a), as the size of CPTs of these are of the order of a)~$1200\times1400$ and b)~$1200\times1400\times700\times643$ respectively. 

\begin{figure}
\centering
\includegraphics[width= 90mm]{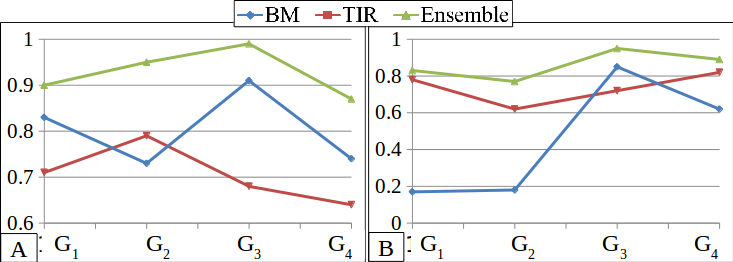}
\caption{x-axis: global characteristics, y-axis: A)~Predictive accuracy for Case-1, B)~Predictive accuracy for Case-2}
\label{fig:case-1}
\end{figure}

Figure \ref{fig:case-1}, shows the prediction accuracy of four global characteristic for Case-1 and Case-2 respectively. Here, the accuracy is a ratio of correctly predicted products to the total number of products. In Case-1, accuracy of Ensemble model is in the range of 85 to 99\%  and it outperforms both SBM and UTS for all four global characteristics.

\textbf{\textit{Baseline Comparison}}: We also compared our approach with record matching method implemented in a framework called FEBRL\cite{christen2008febrl}. For attribute matching, we tried three similarity measures winkler, tokenset, trigam and show the results with winkler which outperforms the rest. We tested this approach for the Case-1 on the smaller dataset (5K products). Table~\ref{table:comparison}, shows the comparison of the prediction accuracy of four global attributes using our Ensemble approach and FEBRL. This suggests that our approach outperforms and also shows that accuracy of FEBRL decreases for high cardinality global attributes. FEBRL did not work on, 26k products, on a machine with 16GB RAM, Intel Core i7-3520M CPU 2.90GHz* 4, 64 bit. We did not try the blocking method as main motive of our problem is to improve accuracy of prediction, and not the time complexity. 
\renewcommand{\arraystretch}{}
\begin{table}
  \centering
  \begin{tabular}{|c|c|c|c|} 
   \hline
   Global att. & Num of states & FEBRL(winkler) & Ensemble \\
   \hline
   $G_1$ & 107 & 86\% & 93\% \\
   \hline
   $G_2$ & 154 & 57\% & 95.2\% \\
   \hline
   $G_3$ & 3 & 99.3\% & 99.2\% \\
   \hline
   $G_4$ & 13 & 95.4\% & 99.4\% \\
   \hline 
  \end{tabular}
\caption{Comparison of our approach with FEBRL}
  \label{table:comparison}
\end{table}
\textbf{Case-2} (Figure \ref{fig:case-1}-B), naturally renders the SBM less accurate, since the training data contains only 10\% of possible states of each global characteristic. However, it is compensated by the performance of UTS, which searches the target set of global attribute values from the
retailer descriptions. Combining these models using our Ensemble model the accuracy of four global characteristics reaches 78 to 93\%.

\begin{figure}[!t]
\centering
\includegraphics[width=100mm]{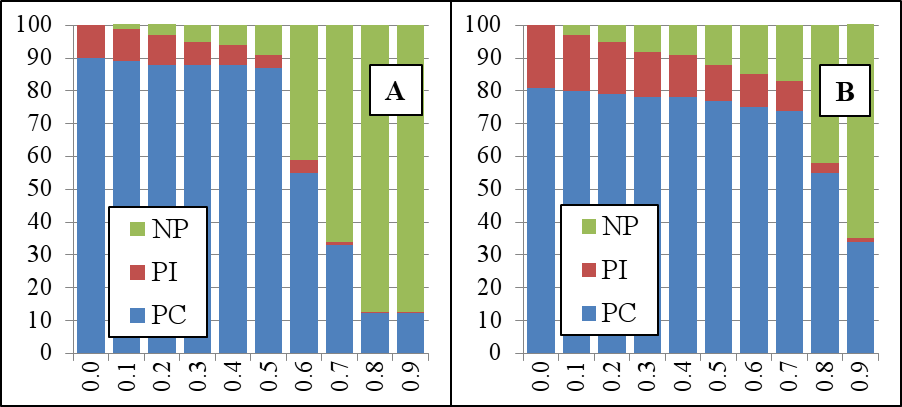}
\caption{\% of Products in each category for different values of $\tau$ on test data for $G_1$ in A)~Case-1 and	 B)~Case-2}
\label{fig:PC-2}
\end{figure}

\textbf{\textit{CoP Threshold for human annotation:}} We define three categories: a)~\textbf{P-C:} Number of products \textit{predicted correctly} by our approach for which CoP $> \tau$. b)~\textbf{P-I:} Number of products \textit{predicted incorrectly}, for which CoP $> \tau$. c)~ \textbf{NP:} Products which we choose \textit{not to predict}, i.e., products with CoP $\leq \tau$. We select $\tau$ in order to maximize P-C and minimize P-I category, while not increasing NP so much that exercise becomes almost entirely manual. Since products in the P-I category are more costly for a company as compared to NP category, we give more weight to P-I while learning $\tau$. Table~\ref{table:categories}, shows the percentage of products in each category (P-C, P-I, NP) on validation set along with the threshold $\tau$ values for both cases. It shows that for given $\tau$, percentage of products in P-C category is in the range of 81-96\% for Case-1, whereas, it ranges from 70 to 96\% for Case-2. Also, the average percentage of products in P-I category is only around 5\%. These numbers establish that CoP is a good measure for reliability of predictions. Figure~\ref{fig:PC-2}, shows the variation in the percentage of products in test set of each category with respect to threshold value $\tau$ for both Case-1 and Case-2, for the global characteristic $G_1$. It validates the optimal values of $\tau$ learned using validation set, 0.5 for Case-1 and 0.6 for Case-2.

\begin{figure}
\centering
\includegraphics[width = 120mm]{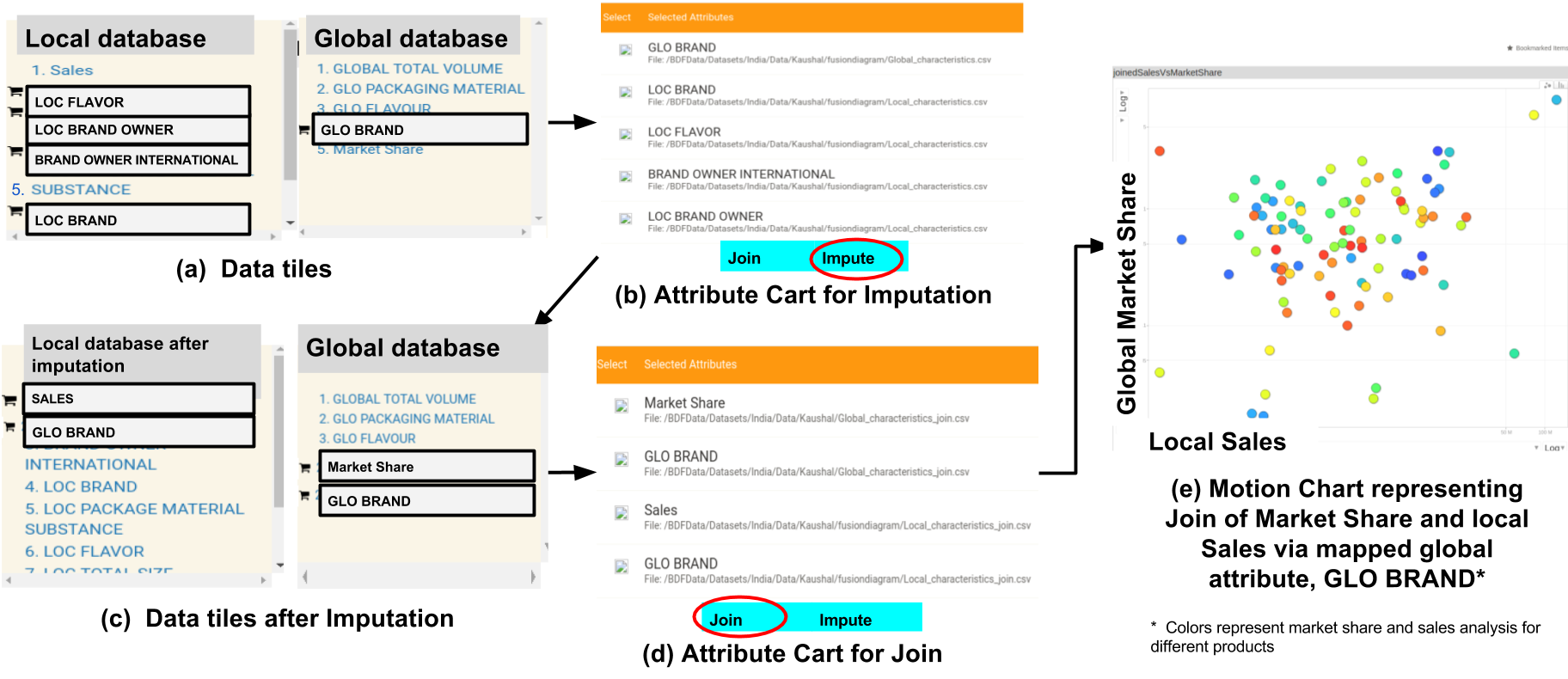}
\caption{Figure showing aggregate analysis of global market share and local sales done using our platform. }
\label{fig:iFuse}
\end{figure}

The process of aggregate analysis, comparing global market share and sales of product categories is carried out in our platform \textbf{\textit{iFuse}}\cite{singh2016visual} (Figure~\ref{fig:iFuse}). Figure~\ref{fig:iFuse}(a),(b) shows the data tile and cart view of iFuse representing the attributes of the local DB and global DB to be linked together. Figure~\ref{fig:iFuse}(c) shows the tile view of the attributes obtained after mapping of local DB to global attribute, here GLO BRAND via ensemble approach, thereby enabling the \textbf{join} of local sales and global market share via common global attribute, GLO BRAND (Figure~\ref{fig:iFuse}(d)). Figure~\ref{fig:iFuse}(e) shows aggregate analysis of different products via motionchart.

\renewcommand{\arraystretch}{}
\begin{table}
  {
  \centering
  \begin{tabular}{|p{0.7cm}|c|c|c|c|c|c|c|c|}
    \hline
    \multirow{2}{*}{Global} &
      \multicolumn{4}{c|}{Case-1} &
      \multicolumn{4}{c|}{Case-2} \\
      \cline{2-9}
    & $\tau$ & P-C & P-I & NP & $\tau$ & P-C & P-I & NP \\
   \hline
   $G_1$ & 0.5 & 92\% & 4\% & 4\% & 0.6 & 82\% & 7\% & 11\% \\
   \hline
   $G_2$ & 0.6 & 81\% & 7\% & 12\% & 0.65 & 74\% & 10\% & 16\% \\
   \hline
   $G_3$ & 0.7 & 96\% & 1\% & 3\% & 0.7 & 96\% & 1\% & 3\% \\
   \hline
   $G_4$ & 0.8 & 86\% & 3\% & 11\% & 0.8 & 85\% & 4\% & 11\% \\
   \hline
  \end{tabular}
  \scriptsize
  \caption{\% of products in each category on Validation set}
  \label{table:categories}
  }
\end{table}


\section{Conclusion}\label{sec:conc}
We have addressed a particular class of record-linkage problems where disparate databases need to be fused in the absence of matching keys
for the limited purpose of aggregate analysis. Our ensemble approach combines supervised Bayesian models with unsupervised textual similarity, 
and also returns confidence along with each prediction. We submit that our approach is likely to be applicable for similar instances of record-linkage in a wide variety of applications, even while attempting to fuse data from external sources, such as social media, sensor data etc.. Such scenarios are becoming increasingly common as the \textit{data lake} paradigm
is gradually replacing the traditional data-warehouse model, driven by the availability
and accessibility of external `big data' sources. 

\bibliographystyle{splncs03}
\bibliography{PAKDD_bib}
\end{document}